\documentclass{sig-alternate-05-2015}
\usepackage{hyperref}
\usepackage{cite}
\usepackage{url}
\usepackage[tight,footnotesize]{subfigure}
\usepackage{balance}
\usepackage{multirow}
\usepackage{fancybox}
\usepackage{mathrsfs}
\usepackage{wrapfig}
\usepackage{graphicx}
\usepackage{amsmath}
\usepackage{mathtools}
\usepackage[mathscr]{euscript}
\usepackage{algorithm}
\usepackage{algorithmic}

\usepackage{amsthm}

\theoremstyle{definition}

\newtheorem{example}{Example}[section]
\theoremstyle{remark}

\graphicspath{{figures/}}
\usepackage{ctable}
\usepackage{bbm}
\usepackage{verbatim}
\usepackage{comment}
\newcommand{\leqnos}{\let\@eqnnum\l@eqnnum}
\newcommand{\reqnos}{\let\@eqnnum\r@eqnnum}

\begin{document}


\CopyrightYear{2016} 
\setcopyright{acmlicensed}
\conferenceinfo{CIKM'16 ,}{October 24 - 28, 2016, Indianapolis, IN, USA}
\isbn{978-1-4503-4073-1/16/10}\acmPrice{\$15.00}
\doi{http://dx.doi.org/10.1145/2983323.2983714}





\clubpenalty=10000 
\widowpenalty = 10000

\title{Bayesian Non-Exhaustive Classification \\ A Case Study: Online Name Disambiguation using Temporal Record Streams
~\titlenote{This research is sponsored by both Mohammad Al Hasan's
NSF CAREER Award (IIS-1149851) and Murat Dundar's
NSF CAREER Award (IIS-1252648). The contents are solely
the responsibility of the authors and do not necessarily represent the official view of NSF.}}

\numberofauthors{3} 

\author{
Baichuan Zhang, Murat Dundar, and Mohammad Al Hasan \\
{Dept. of Computer Science, Indiana University---Purdue University, Indianapolis}\\
{723 W Michigan St., Indianapolis, IN, 46202}\\
{bz3@umail.iu.edu, mdundar@iupui.edu, alhasan@cs.iupui.edu}
}

\maketitle
\begin{abstract}

The name entity disambiguation task aims to partition the records of multiple real-life persons so that
each partition contains records pertaining to a unique person. Most of the existing
solutions for this task operate in a batch mode, where all records to be disambiguated are initially
available to the algorithm. However, more realistic settings require that the name disambiguation task
be performed in an online fashion, in addition to, being able to identify records of new ambiguous entities having no preexisting records. In this work, we propose a Bayesian non-exhaustive classification framework for solving online name disambiguation task. Our proposed method uses a
Dirichlet process prior with a Normal $\times$ Normal $\times$ Inverse Wishart data model which enables identification
of new ambiguous entities who have no records in the training data. For online classification, we use one sweep Gibbs sampler which
is very efficient and effective. As a case study we consider bibliographic data in a temporal stream 
format and disambiguate authors by partitioning their papers into homogeneous groups. Our experimental
results demonstrate that the proposed method is better than existing methods
for performing online name disambiguation task.

\end{abstract}



\section{Introduction}

Name entity disambiguation~\cite{Han.Giles.ea:04, Han.Zha.ea:05,Tang.Fong.ea:12,Zhang.Saha.ea:14} 
(also known as name entity resolution) is an
important problem, which has numerous applications in information retrieval~\cite{Salton.McGill:86, Choudhury.Agarwal.ea:16}, 
digital forensic~\cite{Michaud:01}, 
and social network analysis~\cite{zhang.Choudhury.ea:16, Chen.Zhang.ea:16, Li.Becchi.13}.
In information retrieval domain, name
disambiguation is important for sanitizing search results of ambiguous queries. For example, an online
search query for ``Michael Jordan'' may retrieve pages of former US basketball player, the 
pages of UC Berkeley machine learning professor, and the pages of other persons having that name, 
and name disambiguation is needed to organize those pages in homogeneous groups. In digital forensic,
resolving name ambiguity is essential before inserting a person's profile in a law enforcement database;
failing to do so may cause severe distress to many innocents who are namesakes of a known criminal.
Evidently, name disambiguation plagues the digital library science domain the most. In this domain, 
a key task is to record academic publications in digital repositories and it is 
often the case that the publications of multiple scholars sharing a name are recorded erroneously 
under a unique profile in some repositories. For example, in Google Scholar (GS)\footnote{\url{https://scholar.google.com/}}, which is one of the largest digital libraries for
scholarly publications from various disciplines, there are more than $50$ distinct persons named 
``Wei Wang'', all of whose publications are listed under the same entity. Severe cases of name
ambiguity like this arise in digital library because the first name of an author is 
typically written in an abbreviated form in the citation of many scientific articles. Unresolved
name ambiguity in digital library over- or under-estimates a scholar's citation related impact metrics. 

Due to its importance, the task of name entity disambiguation has allured data mining and database researchers, and over the years, they
have proposed several methods for solving this problem~\cite{Bunescu.Pasca:06,Han.Giles.ea:04}. The
proposed methodologies differ in their learning approaches (supervised~\cite{Han.Giles.ea:04} or unsupervised~\cite{Han.Zha.ea:05}~\cite{Cen.Luo.ea:13}), machine 
learning methodologies (support vector machines~\cite{Han.Giles.ea:04}, Markov random field~\cite{Tang.Fong.ea:12}, graph clustering~\cite{Han.Zha.ea:05}, etc) and the 
data sources that they use (internal data or external data, such as Wikipedia~\cite{Bunescu.Pasca:06}). Many of the proposed
methods are specific to resolving name ambiguity in digital
library, and the major contribution in these works is effective feature engineering involving 
co-authors, publication venues, and research keywords. However, a key limitation of most of the existing methods for name disambiguation is that they operate in a batch mode, where all records to be resolved are
initially accessible to the learning algorithm and a learning model is trained using features extracted
from these records. Hence, they fail to resolve emerging name ambiguities caused from the evolution of
digital data, or they fail to utilize emerging evidences suggestive of merging of name entities which are
separated in the existing state. Re-running a batch learning to catch up with the data evolution is not 
practical due to the enormity of the computation on a large digital repository. So, it is more practical to perform name entity disambiguation task in an incremental fashion by considering the streaming nature of records. We call this {\em online name entity disambiguation}, which is the focus of this work.

\begin{figure}
\includegraphics[width = 3in]{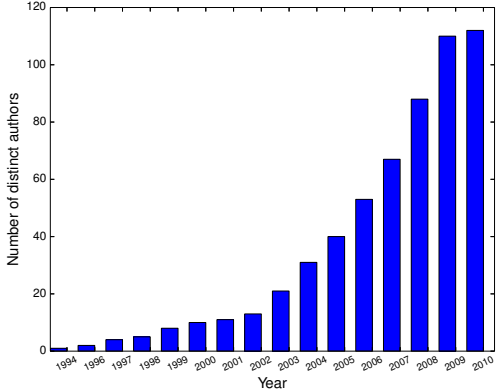}
\caption{Name Ambiguity Evolution for the case of ``Lei Wang"}
\label{fig:intro}
\vspace{-0.2in}
\end{figure}

Perhaps, the most prudent among the existing name disambiguation methodologies is supervised classification 
where for a given name reference~\footnote{By name reference we mean a name string, which may be associated with 
several real-life people entities.} a classification model is trained which classifies each of the records into
a given number of entities (sharing that name). However, such a method is unable to identify records belonging
to emerging name
entities, who do not have any record in the training data. For example, in digital library, every now and
then, papers are authored by a new scholar who is a namesake of some of the existing scholars; as an example,
consider the name reference ``Lei Wang" in Arnetminer~\footnote{\url{http://arnetminer.org}} bibliographic data
repository. In 1999, there were a handful of authors sharing that name, but with each passing year this number has been growing and in 2010 there existed more than
100 authors that have that same name (See Figure~\ref{fig:intro}). A
supervised classification model for the name reference ``Lei Wang'', with a fixed number of classes will never be able to
disambiguate the papers of new authors correctly, as these models have no provision for inclusion of new classes
instantly.

Majority of the existing works on name entity disambiguation consider batch learning solution, however, very
recently online name disambiguation is gaining some traction with a few works~\cite{Qian.Zheng.ea:15,
Khabsa.Giles.ea:15, Ana.Alberto.ea:11, Adriano.Anderson.ea:12}. However, all of these methods
propose a threshold-based approach for identifying emerging classes with a heuristically chosen threshold leading
to unpredictable performance. Real-life online digital library platforms, such as
ResearchGate~\footnote{\url{https://www.researchgate.net/}}, confirm the authorship of ambiguous records with
the authors themselves before including them in their profiles.  However, such a solution may not be very
effective as it relies on a manual verification process which is tedious, and also introduces
significant indexing delays for new records. 

In this work, we solve the online name disambiguation task by a principled approach, namely Bayesian
non-exhaustive classification framework.  We use non-exhaustive learning~\cite{Dundar.Akova.ea:12, Miller.Browning:03}
---a recent development in machine
learning, which considers the scenario that training data may miss 
some classes; it enables our proposed method to disambiguate records belonging
to not only the existing entities, but also the emerging ones. 
Specifically, given a non-exhaustive training dataset, we use a
Dirichlet process prior to model both known and emerging classes, where each class distribution is modeled by a
Normal distribution. We use a common Normal $\times$ Inverse Wishart (NIW) prior to model the mean vectors and
covariance matrices of all class distributions. The hyperparameters of the NIW prior are estimated using data
from known classes facilitating information sharing between known and emerging classes. For a future record, the proposed approach computes class conditional probabilities by considering the possibility that the record may also originate from a new class. Based on these probabilities each record is assigned to one of the existing
classes, or to an emerging class that has no previous records in the training set. We update the set of classes
every time a new class is created by the online model and then use a new classification model to classify
subsequent records. The proposed framework paves the 
way for simultaneous online classification and novel class discovery.

Our contributions can be summarized as follows:

\begin{enumerate}
\item  We study online name disambiguation problem in a non-exhaustive streaming setting and 
propose a self-adjusting Bayesian model that is capable of performing classification, and class discovery, at the same time. 
To the best of our knowledge, our work is the first one to adapt Bayesian non-exhaustive learning for online
name disambiguation task.
\item We propose a one sweep Gibbs sampler to perform online non-exhaustive classification in order to efficiently evaluate the class assignment of an online record.
\item  We demonstrate the utility of our proposed approach on bibliographic datasets and present
experimental results, which demonstrate the superiority of our proposed method over the state-of-the-art methodologies for online name disambiguation.
\end{enumerate}

\section{Related Work}
There exist a large number of works on name entity disambiguation. In terms of methodologies,
supervised~\cite{Bunescu.Pasca:06, Han.Giles.ea:04}, unsupervised~\cite{Han.Zha.ea:05,Cen.Luo.ea:13}, and probabilistic
relational models~\cite{Tang.Fong.ea:12, Wang.Tang.ea:11,Song.Huang.ea:07}, are considered. In a supervised setting, a distinct
entity can be considered as a class, and the objective is to classify each record to one of the classes.
Han et al.~\cite{Han.Giles.ea:04} propose supervised name disambiguation methodologies by utilizing Naive Bayes and 
SVM for name entity disambiguation task. For
unsupervised name disambiguation, the records are partitioned into several clusters with the goal of
obtaining a partition where each cluster contains records from a unique entity. For example, ~\cite{Han.Zha.ea:05}
proposes one of the earliest unsupervised name disambiguation methods for bibliographical data, which is based on $K$-way spectral
clustering. Specifically, they compute Gram matrix representing similarities between different citations and 
apply $K$-way spectral clustering algorithm on the Gram matrix in order to obtain the desired
clusters of the citations. Recently, probabilistic relational models, especially graphical models have also been
considered for the name entity disambiguation task. For instance,~\cite{Tang.Fong.ea:12} proposes to use Markov
Random Fields to address name entity  disambiguation challenge in a unified probabilistic framework. Another work
is presented in ~\cite{Wang.Tang.ea:11} which uses pairwise factor graph model for this task.
~\cite{Zhang.Saha.ea:14, Zhang.Hasan.ea:15, Hermansson.Kerola.ea:13} present approaches for the name disambiguation task on
anonymized graphs and they only leverage graph topological features due to the privacy concerns. 
In addition, ~\cite{Zhang.Dave.ea:15} solves name entity disambiguation problem as the privacy-preserving
classification task such that the anonymized dataset satisfies non-disclosure requirements, at the same time it 
achieves high disambiguation performance.

Most of existing methods above tackle disambiguation task in a batch setting, where all records to be resolved
are initially available to the algorithm, which makes them unsuitable for disambiguating a future record. In
recent years, a few works have considered 
online name entity disambiguation~\cite{Qian.Zheng.ea:15,Khabsa.Giles.ea:15,Adriano.Anderson.ea:12,Ana.Alberto.ea:11, Davis.Veloso.ea:12, Hoffart.Altun.ea:14}, which
perform name disambiguation incrementally without retraining the system every time a new record is received.
Khabsa et al.~\cite{Khabsa.Giles.ea:15} use an online variant of DBSCAN, a well-known density-based clustering
to cluster new records incrementally as they are being added. Since, DBSCAN does not take the number of 
clusters as input, technically speaking, this method is able to adapt to the non-exhaustive scenario simply 
by assigning a new record in a new cluster, as needed. However, DBSCAN is quite susceptible to the choice of
parameter values, and depending on the parameter value, a record belonging to an emerging class can simply
be labeled as an outlier instance.
~\cite{Qian.Zheng.ea:15} proposes a two stage framework for online name disambiguation. The first
stage performs batch name disambiguation to disambiguate all the records no later than a given time threshold using hierarchical agglomerative clustering. The second stage performs
incremental name disambiguation to determine the class membership of a newly added record. However, the method uses a
heuristic threshold to decide on the cluster assignments of new records which makes this approach very susceptible to the choice of threshold parameter.
~\cite{Adriano.Anderson.ea:12} introduces an association rule
based approach for detecting unseen authors in test set.  However, the major drawback of proposed solution is
that it can only identify records of emerging authors in a binary setting but fails to further distinguish among
them. 
~\cite{Davis.Veloso.ea:12} presents the Expectation Maximization based approach for the name
disambiguation in Twitter streaming data. In addition, ~\cite{Hoffart.Altun.ea:14} proposes a threshold based method
to discover emerging entities with ambiguous names in the domain of knowledge base.

Our proposed solution utilizes non-exhaustive learning---a rather recent development in machine learning.
~\cite{Ferit.Dundar.ea:10, Dundar.Akova.ea:12, Miller.Browning:03} are some of the 
existing works related to non-exhaustive learning. Akova et al.~\cite{Ferit.Dundar.ea:10} propose a Bayesian 
approach for detecting emerging classes based on posterior probabilities. However, the decision function 
for identifying emerging classes uses a heuristic threshold and does not consider a prior model over class 
parameters; hence the emerging class detection procedure of this model is purely data-driven. 
Miller et al. ~\cite{Miller.Browning:03} present a mixture model using expectation maximization (EM) for online class discovery. 
~\cite{Dundar.Akova.ea:12} proposes a sequential importance sampling based online inference approach for emerging class
discovery and the work is motivated by a bio-detection application. 


\section{Problem Formulation}~\label{sec:ps}

For a given name reference $a$, assume $X_{n}$ is a stream of records associated with $a$. The subscript
$n$ represents the identifier of the last record in the stream and the value of this identifier increases
as new records are observed in the stream. Each record $x_{i} \in X_{n}$
can be represented by a $d$-dimensional vector which is the feature representation of the record in a 
metric space. In real-life, the name reference $a$ is associated with multiple persons 
(say $k$) all sharing the same name, $a$. The task of name entity disambiguation is to partition 
$X_{n}$ into $k$ disjoint sets such that each partition contains records of a unique person entity.
When $k$ is fixed and known a priori, name entity disambiguation can be solved as a $k$-class classification
task using supervised learning methodologies. However, for many domains the number of classes ($k$)
is not known, rather with new records being inserted in the stream, $X_{n}$ , the number of distinct 
person entities associated with $a$ may increase. The objective of online name entity disambiguation is to 
learn a model that assigns each incoming record into an appropriate partition containing records 
of a unique person entity.

Online name entity disambiguation is marred by several challenges, which we discuss below:

First, for a given record stream $X_{n} = \{x_1, \cdots,  x_{i}, \cdots, x_n\}$, the record $x_i$ 
is classified with the records leading up to $x_{i-1}$, i.e. $X_{i-1}$ is our training data for
this classification task. However, the record $x_i$ may belong to a new person entity (having name $a$) 
with no previous records in $X_{i-1}$. This happens because for online
setting, the number of real-life name entities in $X_{n}$ is not fixed, rather it increases 
over the time. A traditional $k$-class supervised classification model which is trained with records 
of known entities mis-classifies the new emerging record with certainty, leading to an ill-defined classification 
problem. So, for online name entity disambiguation, a learning model is needed which works in non-exhaustive
setting, where instances of some classes are not at all available in the training data. In existing works,
this challenge is resolved using clustering framework where a new cluster is introduced for the emerging 
record of a new person entity,
but this solution is not robust because small changes in clustering parameters make widely varying
clustering outcomes.

The second challenge is that online name entity disambiguation, more often, leads to a severely imbalanced 
classification task. This is due to the fact that in most of the real-life name entity disambiguation problems, 
the size of the true partitions 
of the record set $X_{n}$ follows a power-law distribution. In other words, there are a few persons (dominant 
entities) with the
name reference $a$ to whom the majority of the records belong. Only a few records (typically one or two) belong
to each of the remaining entities (with name reference $a$). Typically, the persons whose records appear at earlier time
are dominant entities, which makes identifying novel entity an even more challenging task.

The third challenge in online name entity disambiguation is related to online learning scenario, where the incoming
record is not merely a test instance of typical supervised learning. Rather, the learning algorithm requires to detect
whether the incoming record belongs to a novel entity, and if so, the algorithm must adapt itself and configure model 
to identify future records of this novel entity. Overall, this requires a 
self-adjusting model that updates the number of classes to accurately classify incoming 
records of both new and existing entities. 

The final challenge in our list is related to temporal ordering of the records. In traditional classification,
records do not have any temporal connotation, so an arbitrary train/test split is permitted. But, for online
setting the model must respect time order of the records, i.e., a future record cannot be used for building a
training model that classifies as older record.

Our proposed model overcomes all the above challenges by using a principled approach.

\section{Entity Disambiguation on Bibliographic Data}

As we have mentioned earlier, name entity disambiguation is a severe issue in digital library domain.  In
many other domains, solving name disambiguation is easier as the method may have access to personalized 
attributes of an entity, 
such as institution affiliation, and email address. But, in digital library, the reference of a paper only includes paper
title, author name, publication venue, and year of publication, which are not sufficient for disambiguation of most of
the name references. Besides, in many citations the first name of the authors are often replaced by initials, which worsen 
the disambiguation issue. As a result, nearly, all the
online bibliographic repositories, including DBLP, Google scholar, ArnetMiner, and PubMed, suffer from this issue.
Nevertheless, these repositories provide timely update of the publication data along with their chronological orders, so they
provide an ideal  for evaluating the effectiveness of an online name entity disambiguation method.

In this work, we use bibliographic data as a case study for online name entity disambiguation. For each name reference $a$,
we build a distinct classification model. The record stream $X_{n}$ for the name reference $a$ is the chronologically ordered stream of scholarly publications where $a$ is one of the authors. To build a feature vector for a paper in
$X_{n}$ we extract features from its author-list, keywords from its paper title, and paper venue (journal/conference). We provide more details of feature construction in the following subsection.

\subsection{Feature Matrix Construction and Preprocessing}~\label{sec:preprocessing}

For a given name reference $a$, say we have a record stream containing $n$ papers for which the name reference 
$a$ is in the
author-list. We represent each paper with a $d$ dimensional feature vector. Then we define a data matrix
$X_{n} \in {\rm I\!R}^{n \times d}$ for $a$, in which each row corresponds to a record (paper) and each
column represents a feature~\footnote{Note that, we use $X_{n}$ to denote both the record stream and record
data matrix.}. In addition, each paper has a class label $l_{i}$ representing 
the $i$-th distinct person entity under name reference $a$, who has authored this paper. Our goal is to
learn a model to partition papers with name reference $a$ in an online setting. 

For a paper, we construct its feature vector (a row of matrix $X_{n}$) using author information, paper title, 
and publication venue as attributes. These features are well-known for name disambiguation in digital
library. For author information, we first aggregate the author-list of all papers into authors, then define a binary feature
representation for each author, indicating his presence or absence in the author-list of that paper. For constructing
keyword based feature, we first filter a set of predefined stop words from the paper titles and use the remaining words
as a feature. For a given paper, we use a binary value based on presence or absence of that word in the title of that
paper. Publication venues are converted to binary feature in the same way. For keywords, instead of using binary value,  we have also considered TF-IDF value, but it does not provide noticeable performance improvement in any of our datasets. A possible reason for that could be we pre-process the data matrix with dimensionality reduction, which is able to capture hidden features directly from a binary data matrix.

Dimensionality reduction step reduces the dimensionality of data matrix $X_{n} \in {\rm I\!R}^{n \times d}$.
This step is important because the matrix $X_{n}$ is severely
sparse with many zero entries. For dimensionality reduction we design an incremental variant of Non-Negative Matrix Factorization (INNMF), which maps $X_{n}$ into a low dimensional space denoted as 
$E_{n} \in {\rm I\!R}^{n \times h}$, where $h$ is the number of hidden dimensions~\footnote{Note that, we use $E_{n}$ to denote both the record stream and record
data matrix after performing dimensionality reduction on $X_{n}$.}. Specifically, we first
perform Non-Negative Matrix Factorization (NNMF)~\cite{Daniel.Seung:01} in the batch mode on the set of training
records initially available. Then in the online mode, we express each sequentially observed record by a linear
combination of the basis vectors generated from initial set of training records, where the coefficients of the basis 
vectors serve as our latent feature values. In order to learn the coefficients, we solve a constrained
quadratic programming problem by minimizing a least square loss function under the constraint that each
coefficient is non-negative. The goal of using INNMF is to discover low dimensional latent features for each
sequentially observed record in an effort to better fit our proposed Normal $\times$ Normal $\times$ Inverse Wishart (NNIW) data model. 

After pre-processing, low dimensional data matrix $E_{n}$ is a collection of time-stamped $n$ record streams with 
which $a$ is associated, namely $E_{n} = \{e_{1}, ..., e_{i}, ..., e_{n}\}$, where $e_{i} \in {\rm I \!R}^{1 \times h}$ is a $h$ dimensional row vector generated from INNMF to represent the $i$-th record
in the given temporal record stream, and all of the $n$ records in $E_{n}$ are sorted temporally, namely  $e_{1}.t \le ... \le e_{i}.t \le ... \le e_{n}.t$, where $e_{i}.t$ 
represents the time for record $e_{i}$.

By using an incessant stream of records, we formalize our online name disambiguation task as follows: given a
time-stamped partition $t_{0}$, we consider two types of records, namely record samples initially available in the
training set with known class membership information, and record samples sequentially observed online with no
verified class membership information. Formally we treat the collection of time-stamped record streams $E_{n} =
\{e_{1},..., e_{i}, ..., e_{n}\}$ as the set of training 
samples initially available, where $e_{1}.t \le ... \le e_{n}.t \le t_{0}$ and $n$ is the total number of samples
in $E_{n}$. As we can see, all of the records in $E_{n}$ occur no later than the given time-stamped threshold $t_{0}$. $Y_{n} = (y_{1}, ..., y_{i}, ..., y_{n})$ is class indicator vector with $y_{i} \in \{l_{1},..,l_{k}\}$, where $k$ is the number of distinct classes in the training set. 
In order to differentiate  records  in the training set from those observed online, we use $\tilde{e}_{i} \in {\rm I \!R}^{1 \times h}$  to represent $i$-th record sequentially observed online. Here we define $\tilde{E}_{r} = \{\tilde{e}_{1}, ..., \tilde{e}_{i}, ..., \tilde{e}_{r} \}$ to be the set of $r$ record samples sequentially available online after time threshold $t_{0}$, 
i.e. $t_{0} \le \tilde{e}_{1}.t \le ... \le \tilde{e}_{r}.t$.
As new ambiguous authors emerge with the incoming records, the set of classes may expand.  Here we denote $\tilde{Y}_{r} = (\tilde{y}_{1}, ..., \tilde{y}_{i}, ..., \tilde{y}_{r} )$ 
to be their corresponding unknown class information with $\tilde{y}_{i} \in \{l_{1}, ..., l_{\tilde{k}_{r}+k}\}$,
where $\tilde{k}_{r}$ is the number of new classes associated with these $r$ records observed online.

Given an arbitrary online record $\tilde{e} \in {\rm I \!R}^{1 \times h}$ at a certain time point,  our proposed Bayesian non-exhaustive classification model computes a probability to decide whether we should assign $\tilde{e}$ to one of the existing classes, or to a new class not yet observed in any of the historical records.

\section{Method}
In this section we discuss Bayesian non-exhaustive name entity disambiguation methodologies. The methodologies discussed  in this section are domain neutral and can be applied to any domain, once an appropriately constructed feature matrix is obtained.

\subsection{Dirichlet Process Prior Model}~\label{sec:dp}

We model the set of record streams $E_{n} = \{e_{1},...,e_{n}\}$ by using a set of latent parameters 
$\{\theta_{1},...,\theta_{n}\}$. Each $\theta_{i}$ is drawn independently and identically (iid) by a Dirichlet Process~\cite{Ferguson:73} (DP), 
while each record $e_{i} \in {\rm I \!R}^{1 \times h}$ is distributed according 
to an unknown distribution $F(\theta_{i})$ parametrized by $\theta_{i}$. Mathematically,

\begin{eqnarray}
\label{eq:dp}
G &\sim& DP(\alpha, H) \nonumber \\
\theta_{i} &\sim& G \nonumber \\
e_{i}|\theta_{i} &\sim& F(\theta_{i}) \nonumber \\
\end{eqnarray}

where $G$ is a random discrete probability measure defined by a base distribution $H$ along with
the precision parameter $\alpha > 0$. 

It is a common practice to use stick breaking construction~\cite{Sethuraman:94} to represent samples drawn from a Dirichlet process as below:

\begin{eqnarray}
\label{eq:sbc}
\phi_{i} &\sim& H \nonumber \\
\beta_{i} &\sim& Beta(1, \alpha) \nonumber \\
\pi_{i} &=& \beta_{i}\prod_{j=1}^{i-1}(1-\beta_{j}) \nonumber \\
\end{eqnarray}

As shown in Equation~\ref{eq:sbc}, in order to simulate the process of stick breaking construction, imagine we have a stick of length $1$ to represent total probability. 
We first generate each point $\phi_{i}$ from base distribution $H$, which originates from our proposed NNIW data model (Details in Section~\ref{sec:datamodel}).  
Then we sample a random variable $\beta_{i}$ from Beta(1, $\alpha$) distribution.  
After that we break off a fraction $\beta_{i}$ of the remaining stick as the weight of parameter $\phi_{i}$, denoted as $\pi_{i}$. 
In this way it allows us to represent random discrete probability measure $G$ as a probability mass function in terms of infinitely many 
$\phi_{1}, ..., \phi_{\infty}$ and their corresponding weights  $\pi_{1}, ..., \pi_{\infty}$ 
yielding $G = \displaystyle \sum_{i=1}^{\infty}\pi_{i}\delta_{{\phi_{i}}}$, where $\delta_{{\phi_{i}}}$ is the point mass of $\phi_{i}$.

\subsection{Bayesian Non-Exhaustive Online Classification}\label{sec:bnl}

Consider that at a certain time point, we have a set of $n$ training records for name reference $a$,  denoted by $E_{n} = \{e_{1}, ..., e_{n}\}$ where each record is assigned to a latent parameter from the set  $\{\theta_{1},...,\theta_{n}\}$. A future online record $\tilde{e}$ is assigned to $\tilde{\theta}$ with a probability 
\begin{equation}
\label{eq:condp}
\tilde{\theta} | \theta_{1}, ..., \theta_{n} \sim \frac{\alpha}{\alpha+n}H + \frac{1}{\alpha + n} \sum_{i=1}^{n} \delta_{\theta_{i}} 
\end{equation}

The conditional prior distribution in Equation~\ref{eq:condp} can be interpreted by a mixture of two distributions. Specifically, given
the future record $\tilde{e}$ with parameter $\tilde{\theta}$, $\tilde{e}$ belongs to the base distribution $H$ with 
probability $\frac{\alpha}{\alpha + n}$ and it originates from random discrete probability measure $G$ generated 
from DP with probability $\frac{n}{\alpha + n}$.  Since $G$ is a discrete distribution, each record in 
a sequence of $n$ records generated from $G$ may not belong to a distinct $\theta_i$.  
If we assume that there are $k$ distinct values of $\theta$ associated with the first $n$ records, then we can re-write Equation~\ref{eq:condp} as below:

\begin{equation}
\label{eq:condp2}
\tilde{\theta}|\theta_{1}, ..., \theta_{n} \sim \frac{\alpha}{\alpha+n}H + \frac{1}{\alpha + n} \sum_{j=1}^{k} n_{j}\delta_{\theta_{j}} 
\end{equation} 
where in the last expression of Equation~\ref{eq:condp2}, $\theta_{j}$ represents each distinct $\theta$, and $n_{j}$ denotes the number of times $\theta_{j}$ appears among the sequence of $n$ records.
Furthermore, each $\theta_{j}$ is associated with a unique class $l_{j}$ whose corresponding record $e_{j}$ is generated according to $F(\theta_{j})$. 
Thus after a sequence of $n$ records are observed, $\tilde{y}$, the class membership of the future record $\tilde{e}$, 
is equal to $l_{j}$ with probability $\frac{n_{j}}{\alpha + n}$, and it is equal to $l_{k+1}$ with probability 
$\frac{\alpha}{\alpha+n}$, where $l_{k+1}$ is an emerging class which has not been observed in the sequence of $n$ 
training records.  

Next we incorporate the data model into the Dirichlet process prior and use the conditional posterior to determine whether $\tilde{e}$ should be assigned to one of the existing classes or to an emerging class. More specifically, we are interested to evaluate the following distribution:

\begin{eqnarray}
\label{eq:5}
&&p(\tilde{\theta} |\tilde{e}, \theta_{1},...,\theta_{n}) \nonumber \\
&\propto& \frac{\alpha}{\alpha+n}p(\tilde{e})p(\tilde{\theta}|\tilde{e}) \nonumber \\
&+& \frac{1}{\alpha+n}\displaystyle\sum_{j=1}^{k}n_{j}p(\tilde{e}|\theta_{j})\delta_{\theta_{j}} \nonumber \\   
\end{eqnarray}
As we can observe from Equation~\ref{eq:5}, $\tilde{e}$ either belongs to a new class $l_{k+1}$ with the probability
proportional to $\frac{\alpha}{\alpha+n}p(\tilde{e})$, or to a existing class $l_{j}$ with the probability proportional to
$\frac{n_{j}}{\alpha+n}p(\tilde{e}|\theta_{j})$.

Due to the fact that $\theta_{j}$ is not known for online record sequence, 
here we replace $p(\tilde{e}|\theta_{j})$ with the conditional predictive distribution
$p(\tilde{e}|D_{j})$, where $D_{j}=\{e_{i}\}_{i \in l_{j}}$ represents the subset of records with class label $l_{j}$. 
Thus for the online record $\tilde{e}$, given the class membership information of all of $n$ sequence of records processed before $\tilde{e}$,
the following decision making function $g(\tilde{e})$ decides whether $\tilde{e}$ belongs to an emerging class or to one 
of the existing ones.

\begin{equation}
\label{eq:df1}
     g(\tilde{e}) = 
\begin{cases}
      \tilde{y} = l_{j^{\ast}}, & \text{if } \\
      \frac{n_{j^{\ast}}}{\alpha + n }p(\tilde{e}|D_{j^{\ast}})  \geq \frac{\alpha}{\alpha + n}p(\tilde{e}) \\
      \tilde{y} = l_{k+1}, & \text{if } \\ 
      \frac{n_{j^{\ast}}}{\alpha + n}p(\tilde{e} | D_{j^{\ast}})  < \frac{\alpha}{\alpha + n}p(\tilde{e}) 
\end{cases}
\end{equation}

where $j^{\ast} = argmax_{j}\left \{\frac{n_{j}}{\alpha + n}p(\tilde{e} | D_{j})\right \}_{j=1}^{k}$.

\subsection{Gibbs Sampler for Non-Exhaustive Learning}~\label{sec:oi}

\begin{figure}
\includegraphics[width = 3.3in]{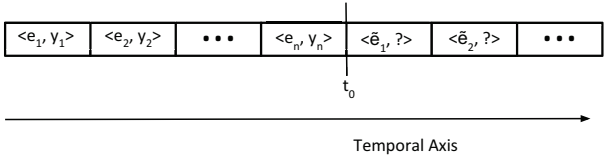}
\caption{Demonstration of Temporal Record Stream}
\label{fig:temporal}
\vspace{-0.1in}
\end{figure}

As shown in Figure~\ref{fig:temporal}, since we only have access to the class membership information $Y_{n}$ for the streaming records initially available in the training set $E_{n}$, which occur no later than the given time-stamped threshold $t_{0}$, 
and the class membership $\tilde{Y}_{i-1} = (\tilde{y}_1, \cdots, \tilde {y}_{i-1})$ of online sequentially observed record stream $\tilde{E}_{i-1} = \{\tilde{e}_1, \cdots, \tilde {e}_{i-1}\}$ is not verified,  thus it is impractical to evaluate
the decision function shown in Equation~\ref{eq:df1} directly. Instead in order to respect the temporal order of each record, we utilize one sweep Gibbs sampler for online 
classification to efficiently evaluate the probability of an online record belonging to an emerging class or 
to one of the existing ones. If $\tilde{y}_i$ is the predicted class label of $i$-th online observed record $\tilde{e}_i$, the conditional distributions of the
class indicator variable $\tilde{y}_{i}$ can be easily obtained via one sweep Gibbs sampler. Specifically, we are interested to sample from the following conditional distribution:

\begin{eqnarray}
\label{eq:2}
&& p(\tilde{y}_{i} | \tilde{E}_{i-1}, \tilde{Y}_{i-1}, E_{n}, Y_{n}) \nonumber \\
&=&
\begin{cases}
      l_{j}   \text{ with probability $\propto$} 
      \frac{n_{j}}{\alpha + n + i -1}p(\tilde{e}_{i} | D_{j})  \\
      l_{k+\tilde{k}_{i-1} + 1}   \text{ with probability $\propto$ }  
      \frac{\alpha}{\alpha + n + i - 1}p(\tilde{e}_{i})  \\
\end{cases}      
\end{eqnarray} 
where $j \in \{1, ..., k+\tilde{k}_{i-1}\}$, 
and $\tilde{k}_{i-1}$ is the number of new classes among the first $(i-1)$ online records.

\subsection{Data Model}\label{sec:datamodel}

The one-sweep Gibbs sampling based online classification technique shown in Equation~\ref{eq:2}
requires the evaluation of the conditional predictive distribution $p(\tilde{e}_{i} | D_{j})$ and marginal distribution $p(\tilde{e}_{i})$.
Fortunately a closed-form solution for both distributions does exist for the Normal $\times$ Normal $\times$ Inverse Wishart (NNIW) model.

We assume that the collected streaming records $E_{n} = \{e_{1},..., e_{i}, ..., e_{n}\} \in {\rm I\!R}^{n \times h}$ 
generated by the INNMF step has the property of unimodality as each record is expressed by a linear combination of the basis vectors and the coefficients of the basis vectors are used as our features. Thus we use  a normally distributed data model, which can model class distributions fairly well. We present our data model under the Bayesian non-exhaustive classification framework as below: 

\begin{eqnarray}
\label{eq:4}
&E_{n} \sim \mathcal{N}(\mathbf{\mu}, \Sigma)& \nonumber \\
&\mathbf{\mu} \sim \mathcal{N}(\mu_0, \kappa^{-1}\Sigma) & \nonumber \\
&\Sigma \sim W^{-1}(\Sigma_0, m) & \nonumber \\
\end{eqnarray}

where $\mu_{0}$ is the prior mean and $\kappa$ is a scaling constant
that controls the deviation of the class conditional
mean vectors from the prior mean. The smaller
$\kappa$ is, the larger the between class scattering will be.
The parameter $\Sigma_{0}$ is a positive definite matrix that
encodes our prior belief about the expected $\Sigma$. The
parameter $m$ is a scalar that is negatively correlated
with the degrees of freedom. In other words the larger
$m$ is, the less $\Sigma$ will deviate from $\Sigma_{0}$ and vice versa.

In addition, the base distribution $H$ in our proposed Dirichlet process prior model 
originates from the joint conjugate distribution $p(\mu, \Sigma)$, which is defined as below:

\begin{eqnarray}
H &=& p(\mu, \Sigma) \nonumber \\
&=& p(\mu|\Sigma)\times p(\Sigma) \nonumber \\
&=& \mathcal{N}(\mu|\mu_{0}, \frac{\Sigma}{\kappa}) \times W^{-1}(\Sigma|\Sigma_{0}, m) \nonumber \\
\end{eqnarray}

Here we define the sample mean of these $n$ streaming records to be $\overline{\mu} = \frac{1}{n}\displaystyle \sum_{i=1}^{n}e_{i}$, and
sample covariance matrix $S = \frac{1}{n-1}\displaystyle \sum_{i=1}^{n}(e_{i} - \overline{\mu})(e_{i} - \overline{\mu})^{T}$.
As sample mean $\overline{\mu}$ and sample covariance matrix $S$ are sufficient statistics for the multivariate 
normally distributed data, in order to compute the conditional predictive distribution $p(\tilde{e}_{i}|D_{j})$, we can replace it with $p(\tilde{e}_{i}|\overline{\mu}_{j}, S_{j})$,
where $\overline{\mu}_{j}$ and $S_{j}$ are sample mean and sample covariance matrix for class $l_{j}$. The mathematical 
derivation of $p(\tilde{e}_{i}|\overline{\mu}_{j}, S_{j})$ is available in~\cite{Anderson:84} and the result is in the form of a Multivariate Student-t distribution as is shown below:

\begin{eqnarray}
\label{eq:10}
&p(\tilde{e}_{i} | \overline{\mu}_{j}, S_{j}) = stu-t(\mu_{s}, \Sigma_{s}, df_{s})& \nonumber \\
\label{eq:stud-t}
&\mu_{s} = \frac{n_{j}\overline{\mu}_{j} + \kappa\mu_0}{n_{j}+\kappa}& \nonumber \\
& \scalebox{0.85}{$\Sigma_{s} = \frac{n_{j}+\kappa+1}{(n_{j}+\kappa)(n_{j}+m+1-h)}\left(\Sigma_0 + (n_{j}-1)S_{j}+\frac{n_{j}\kappa}{n_{j}+\kappa} (\mu_{0} - \overline{\mu}_{j}) (\mu_{0}- \overline{\mu}_{j})^T\right)$} & \nonumber \\ 
&df_{s} = n_{j}+m+1-h & \nonumber \\ 
\end{eqnarray}
where $\mu_{s}$ is a $h \times 1$ mean vector, $\Sigma_{s}$ is a $h \times h$ scale matrix, and $df_{s}$ is the degree of freedom of the obtained
Multivariate Student-t distribution.

Besides, we can set $D_{j} = \emptyset$ in $p(\tilde{e}_{i} |D_{j})$ in order to evaluate the marginal distribution $p(\tilde{e}_{i})$ 
and it becomes another Multivariate Student-t distribution. Therefore we could use Equation~\ref{eq:10} to evaluate 
both conditional predictive distribution $p(\tilde{e}_i|D_j)$ and marginal distribution $p(\tilde{e}_i)$ for one sweep Gibbs sampler shown in Equation~\ref{eq:2}.

\begin{table}[t!]
\centering
\scalebox{0.80}{
\begin{tabular}{c c c c}
\toprule
Name Reference & \# Records & \# Attributes & \# Distinct Authors  \\ \midrule
Kai Zhang & 66 & 480 & 24 \\
Bo Liu & 124 & 739 & 47 \\
Jing Zhang & 231 & 1440 & 85 \\
Yong Chen & 84 & 545 & 25 \\
Yu Zhang & 235 & 1427 & 72 \\
Hao Wang & 178 & 1058 & 48 \\
Wei Xu & 153 & 1023 & 48 \\
Lei Wang & 308 & 1797 & 112 \\
Bin Li & 181 & 1131 & 60 \\
Feng Liu & 149 & 919 & 32 \\
Lei Chen & 196 & 1052 & 40 \\
Ning Zhang & 127 & 740 & 33 \\
David Brown & 61 & 437 & 25 \\
Yang Wang & 195 & 1227 & 55 \\
Gang Chen & 178 & 1049 & 47 \\
\bottomrule
\end{tabular}}
\caption{Arnetminer name disambiguation dataset}
\label{tab:dataset}
\vspace{-0.20in}
\end{table}

\section{Experiments and Results}

In this section, we discuss our experimental results which validate the performance of our proposed Bayesian non-exhaustive classification method for online name entity disambiguation on real-life data. This results also demonstrate the superiority 
of our proposed method against current state-of-the-art in online name entity disambiguation. 

\subsection{Experimental Setting}

We use Arnetminer's name entity disambiguation dataset~\footnote{\url{https://aminer.org/disambiguation}} for our experiment. 
This dataset contains author records of $15$ highly ambiguous name references selected from Arnetminer database. The name references are shown in Table~\ref{tab:dataset}. In this
table, for each name reference we also show the number of records, the number of binary attributes (explained in Section~\ref{sec:preprocessing}), and the number of distinct authors associated with that name reference.

For each of the 15 name references in the above dataset, we train a separate model for the disambiguation task.
To simulate streaming data, we sort the records (papers) in temporal order and make train-test partition.
Specifically, we put the publication records of most recent $N$ years into the test set and the papers from
earlier years in the training set. Then we measure the performance of our proposed model by varying the value of $N$.
For a given train-test partition, we first train the model using the training set initially available, then we 
process the records in the test set one-by-one in order to simulate streaming data in the online setting.
For evaluation metric, we use macro-F1 measure, which is average of F1-measure of each class. The range of
macro-F1 measure is between $0$ and $1$, and a higher value indicates better disambiguation performance. 

Our proposed method has the following tunable parameters, which we tune by using the training data. The first
among those is the latent dimensionality $h$ for INNMF (Section~\ref{sec:preprocessing}). We consider different
values between $5$ and $20$ and set the value that achieves the best macro-F1 measure on the training set by 
cross validation. The second parameter is $\alpha$ in the Dirichlet process prior model (Section~\ref{sec:dp}), which 
controls the probability of assigning an incoming record to a new class entity. It plays a critical role in the number 
of generated classes in the name disambiguation process. 
In this work, in order to estimate the parameter $\alpha$,
we first encode our prior belief about the odds of encountering a new class by a prior probability value $p$, which can be set by measuring the probability of emerging records in a temporal sub-partition of the training data. Given this prior, we estimate $\alpha$ by
empirical Bayes through collecting a large number of samples from a Chinese Restaurant Process (CRP)~\cite{Aldous:85} for varying 
values of $\alpha$ and then choosing the value that minimizes the difference between the empirical and true value of $p$.
Our final set of parameters are the prior distribution of NIW model: $(\Sigma_{0}, m, \mu_{0}, \kappa)$ (see Section~\ref{sec:datamodel}). We estimate these offline by using data records in the training set. Specifically, we use 
the mean of the training set to estimate $\mu_{0}$, and set $\Sigma_{0}$ as the pooled covariance $S_{p}$ as suggested in~\cite{Greene.William:89}. Here, $S_{p}$ is defined as below: 

\begin{small}
\begin{equation}
S_{p} = \frac{(m-h-1)\displaystyle\sum_{j=1}^{k}(n_{j}-1)S_{j}}{n-k}
\end{equation}
\end{small} 
%
%
where $h$ is the number of latent dimension from NNMF step, $n_{j}$ and $S_{j}$ are
the number of samples and sample covariance matrix with respect to class $l_{j}$ in the training set. Besides,  we coarsely tune $m$ and $\kappa$ with three values each, namely  $\kappa = 10, 50 , 100$ and $m = h + 10, h + 50, h + 100$ and select the parameter combination with best disambiguation performance  by cross validation on the training set.

In order to illustrate the merit of our proposed approach, we compare our model with the following benchmark techniques. Among
these the first two are existing state-of-the-art online name entity disambiguation methods, and the latter two are baselines that
we have designed.

\begin{enumerate}

\item \textbf{Qian's Method~\cite{Qian.Zheng.ea:15}} 
Given the collection of training records initially available, for a new record, Qian's method computes class conditional probabilities for existing classes. This approach assumes that all the attributes are independent and the procedure of probability computation is based on the occurrence count of each attribute in all records of each class. 
Then the computed probability is compared with a pre-defined threshold value to determine whether the newly added record should be assigned to an existing class, or to a new class not yet included in the previous data. 

\item \textbf{Khabsa's Method~\cite{Khabsa.Giles.ea:15}}
Given the collection of training records initially available this approach first computes the $\epsilon$-neighborhood density for each online sequentially observed record. The $\epsilon$-neighborhood density of a new record is considered as the set of records within $\epsilon$ euclidean distance from that record. Then if the neighborhood is sparse, the new record is assigned to a new class. Otherwise, it is classified  into the existing class that contains the most records in the $\epsilon$-neighborhood of the new record.  

\item \textbf{BernouNaive-HAC:} In this baseline, we first model the data with a multivariate Bernoulli distribution (features are binary, so Bernoulli distribution is used) and train a Naive Bayes classifier. This classifier returns class conditional probabilities for each record in the test set which we use as meta features in a hierarchical agglomerative clustering (HAC) framework.

\item \textbf{NNMF-SVM-HAC:} We perform NNMF on our binary feature matrix and use the coefficients returned by NNMF to train a linear SVM. Class conditional probabilities for each test record are used as meta features in a hierarchical agglomerative clustering (HAC) framework the same way described above.
\end{enumerate}

\begin{table*}[t!]
\centering
\scalebox{0.73}{
\begin{tabular}{c| c c c c| c c c c c}
\toprule
Name & \# train & \# test & \# emerging & \# emerging & {\bf Our Method} & BernouNaive- & NNMF- & Qian's & Khabsa's \\
Reference  & records          & records     & records    & classes  &  & HAC & SVM-HAC& Method~\cite{Qian.Zheng.ea:15} & Method~\cite{Khabsa.Giles.ea:15} \\
\midrule
Kai Zhang & 42 & 24 & 15 & 8 & \textbf{0.683 (0.041)} & 0.605 & 0.621 & 0.619 & 0.518 \\
Bo Liu & 99 & 25 & 11 & 8 & \textbf{0.786 (0.033)} & 0.733 & 0.719 & 0.714 & 0.559 \\
Jing Zhang & 121 & 110 & 56 & 35 & \textbf{0.691 (0.028)} & 0.554 & 0.566 & 0.590 & 0.631 \\
Yong Chen & 70 & 14 & 5 & 5 & \textbf{0.889 (0.016)} & 0.852 & 0.794 & 0.848 & 0.833 \\
Yu Zhang & 124 & 111 & 62 & 30 & \textbf{0.535 (0.013)} & 0.498 & 0.516 & 0.515 & 0.502 \\
Hao Wang & 148 & 30 & 9 & 8 & \textbf{0.747 (0.026)} & 0.635 & 0.639 & 0.702 & 0.581 \\
Wei Xu & 127 & 26 & 11 & 10 & \textbf{0.844 (0.033)} & 0.811 & 0.750 & 0.767 & 0.689 \\
Lei Wang & 245 & 63 & 28 & 24 & \textbf{0.755 (0.012)} & 0.701 & 0.708 & 0.703 & 0.620 \\
Bin Li & 154 & 27 & 11 & 9 & \textbf{0.805 (0.029)} & 0.775 & 0.733 & 0.775 & 0.743 \\
Feng Liu & 104 & 45 & 6 & 5 & \textbf{0.579 (0.031)} & 0.501 & 0.499 & 0.399 & 0.339 \\
Lei Chen & 96 & 100 & 24 & 18 & 0.356 (0.043) & \textbf{0.646} & 0.527 & 0.430 & 0.222 \\
Ning Zhang & 97 & 30 & 16 & 12 & 0.635 (0.021) & 0.669 & \textbf{0.685} & 0.647 & 0.608 \\
David Brown & 48 & 13 & 4 & 3 & 0.833 (0.019) & \textbf{0.902} & 0.593 & 0.816 & 0.450 \\
Yang Wang & 118 & 77 & 38 & 20 & 0.449 (0.033) & 0.513 & \textbf{0.546} & 0.315 & 0.440 \\
Gang Chen & 113 & 65 & 20 & 14 & \textbf{0.821 (0.004)} & 0.474 & 0.467 & 0.401 & 0.357 \\
\bottomrule
\end{tabular}}
\caption{Comparison of Macro-F1 values between the proposed method and four other competing methods using records with most recent 2 years as test set}
\label{tab:result1}
\vspace{-0.10in}
\end{table*}

For all the competing methods, we use identical set of features (before dimensionality reduction). 
We vary the probability 
threshold value of Qian's method and $\epsilon$ value of Khabsa's method by cross validation on the
training dataset.
and select the ones that obtain the best disambiguation performance in terms of macro-F1 score.  
For BernouNaive-HAC and NNMF-SVM-HAC methods, during the hierarchical agglomerative clustering step, 
we tune the number of 
clusters in training set by cross validation in order to get the best disambiguation result. 

For both data processing and model implementation, we implement our own code in Python and use
NumPy~\footnote{\url{http://www.numpy.org/}}, SciPy~\footnote{\url{https://www.scipy.org/}},
CVXOPT~\footnote{\url{http://cvxopt.org/}} and scikit-learn~\footnote{\url{http://scikit-learn.org/stable/}} libraries for linear algebra, optimization and machine learning operations. 
We run all the experiments on a 2.1 GHz Machine 
with 4GB memory running Linux operating system.

\begin{table*}[t!]
\centering
\scalebox{0.73}{
\begin{tabular}{c| c c c c| c c c c c}
\toprule
Name & \# train & \# test & \# emerging & \# emerging & {\bf Our Method} & BernouNaive- & NNMF- & Qian's & Khabsa's \\
Reference  & records          & records     & records    & classes  &  & HAC & SVM-HAC& Method~\cite{Qian.Zheng.ea:15} & Method~\cite{Khabsa.Giles.ea:15} \\
\midrule
Kai Zhang & 27 & 39 & 20 & 10 & \textbf{0.602 (0.021)} & 0.503 & 0.584 & 0.520 & 0.510 \\
Bo Liu & 66 & 58 & 29 &  21 & \textbf{0.759 (0.011)} & 0.612 & 0.606 & 0.612 & 0.631 \\
Jing Zhang & 82 & 149 & 77 & 47 & \textbf{0.618 (0.022)} & 0.480 & 0.506 & 0.523 & 0.419  \\
Yong Chen & 54 & 30 & 12 & 8 & \textbf{0.865 (0.047)} & 0.615 & 0.701 & 0.615 & 0.545 \\
Yu Zhang & 87 & 148 & 71 & 38 & 0.457 (0.013) & 0.445 & \textbf{0.615} & 0.447 & 0.412 \\
Hao Wang & 115 & 63 & 17 & 12 & \textbf{0.698 (0.031)} & 0.513 & 0.572 & 0.540 & 0.512 \\
Wei Xu & 101 & 52 & 17 & 14 & \textbf{0.734 (0.051)} & 0.683 & 0.603 & 0.635 & 0.586 \\
Lei Wang & 173 & 135 & 67 & 45 & \textbf{0.723 (0.044)} & 0.560 & 0.522 & 0.536 & 0.428 \\
Bin Li & 108 & 73 & 37 & 23 & \textbf{0.777 (0.009)} & 0.532 & 0.574 & 0.588 & 0.545 \\
Feng Liu & 70 & 79 & 9 & 8 & \textbf{0.544 (0.017)} & 0.488  & 0.527 & 0.379 & 0.424 \\
Lei Chen & 65 & 131 & 39 & 25 & 0.332 (0.029) & \textbf{0.493} & 0.447 & 0.398 & 0.176 \\
Ning Zhang & 76 & 51 & 32 & 19 & 0.589 (0.034) & \textbf{0.744} & 0.531 & 0.420 & 0.378 \\
David Brown & 39 & 22 & 17 & 7 & 0.734 (0.008) & 0.751 & 0.631 & \textbf{0.752} & 0.478 \\
Yang Wang & 92 & 103 & 46 & 25 & \textbf{0.436 (0.012)} & 0.313 & 0.298 & 0.225 & 0.240 \\
Gang Chen & 89 & 89 & 27 & 19 & \textbf{0.799 (0.008)} & 0.347 & 0.407 & 0.383 & 0.221 \\
\bottomrule
\end{tabular}}
\caption{Comparison of Macro-F1 values between the proposed method and four other competing methods using records with most recent 3 years as test set}
\label{tab:result2}
\vspace{-0.10in}
\end{table*}

\begin{table*}[t!]
\centering
\scalebox{0.73}{
\begin{tabular}{c| c c c c| c c c c c}
\toprule
Name & \# train & \# test & \# emerging & \# emerging & {\bf Our Method} & BernouNaive- & NNMF- & Qian's & Khabsa's \\
Reference     & records          & records     & records    & classes  &  & HAC & SVM-HAC& Method~\cite{Qian.Zheng.ea:15} & Method~\cite{Khabsa.Giles.ea:15} \\
\midrule
Kai Zhang & 12 & 54 & 39 & 17 & \textbf{0.523 (0.017)} & 0.448 & 0.471 & 0.506 & 0.469 \\
Bo Liu & 42 & 82 & 40 & 24 & 0.480 (0.023) & 0.439 & \textbf{0.540} & 0.497 & 0.414 \\
Jing Zhang & 53 & 178 & 105 & 60 & \textbf{0.502 (0.018)} & 0.455 & 0.366 & 0.407 & 0.413 \\
Yong Chen & 46 & 38 & 15 & 10 & \textbf{0.669 (0.039)} & 0.588 & 0.617 & 0.605 & 0.303 \\
Yu Zhang & 51 & 184 & 119 & 53 & \textbf{0.410 (0.009)} & 0.401 & 0.398 & 0.315 & 0.302 \\
Hao Wang & 90 & 88 & 26 & 19 & \textbf{0.649 (0.028)} & 0.454 & 0.508 & 0.521 & 0.433 \\
Wei Xu & 76 & 77 & 32 & 21 & \textbf{0.695 (0.049)} & 0.412 & 0.437 & 0.525 & 0.507 \\
Lei Wang & 131 & 177 & 86 & 59 & \textbf{0.595 (0.035)} & 0.502 & 0.558 & 0.498 & 0.383 \\
Bin Li & 73 & 108 & 64 & 33 & \textbf{0.625 (0.021)} & 0.444 & 0.470 & 0.439 & 0.502 \\
Feng Liu & 46 & 103 & 36 & 14 & \textbf{0.395 (0.014)} & 0.378 & 0.385 & 0.321 & 0.295 \\
Lei Chen & 38 & 158 & 56 & 29 & 0.302 (0.033) & \textbf{0.453} & 0.416 & 0.234 & 0.190 \\
Ning Zhang & 65 & 62 & 33 & 20 & \textbf{0.547 (0.024)} & 0.531 & 0.474 & 0.412 & 0.415 \\
David Brown & 35 & 26 & 22 & 11 & \textbf{0.707 (0.006)} & 0.662 & 0.677 & 0.456 & 0.417 \\
Yang Wang & 68 & 127 & 64 & 33 & 0.474 (0.023) & \textbf{0.476} & 0.457 & 0.121 & 0.361 \\
Gang Chen & 61 & 117 & 38 & 25 & \textbf{0.646 (0.015)} & 0.307 & 0.405 & 0.315 & 0.084 \\
\bottomrule
\end{tabular}}
\caption{Comparison of Macro-F1 values between the proposed method and four other competing methods using records with most recent 4 years as test set}
\label{tab:result3}
\vspace{-0.10in}
\end{table*}

\subsection{Evaluation of Various Name Disambiguation Methods}

We vary the train/test split to observe the performance comparison between
our proposed method and other competing methods for different experimental settings.
In our first setting, records from the latest two years are used in the test split,
and the remaining records are used in the train split. In two other settings,
records from the latest three and latest four years are used in test split, respectively.
Table~\ref{tab:result1}, Table~\ref{tab:result2},
and Table~\ref{tab:result3} show the experimental results for these three settings.
In all these three tables, the rows correspond to the fifteen name references.
The last five columns show the performance of entity disambiguation of the corresponding
name reference using macro-F1 score. Since one sweep Gibbs sampler in our proposed model 
is a randomized method,  for each name reference we execute the method $20$ times and 
report the average macro-F1 score. For our method, we also show the standard deviation in the 
parenthesis~\footnote{Standard deviation for other methods are not shown due to the fact that they are not randomized.}.
For better visual comparison, we highlight the best macro-F1 score of each name reference
with boldface font. 

The ``\#train records'' and ``\#test records'' columns in these tables represent the  
number of training and test records; ``\#emerging records'' is the number of records in test 
set with their corresponding classes not represented in the initial training set, and 
``\#emerging classes'' denotes the number of emerging classes not represented in the training set.
For all rows, as we compare the values in these columns across Table~\ref{tab:result1}, Table~\ref{tab:result2}, and Table~\ref{tab:result3}, the number of training records decreases,
the number of test records, emerging records, and emerging classes increase. It makes
the disambiguation task in the first setting (2 years test split) the easiest, and the
third setting (4 years of test split) the hardest. This is reflected in the macro-F1
values of all the name references across these three tables. For example, for the first name reference,
Kai Zhang, macro-F1 score of our method across these three tables are 0.683, 0.602, and 0.523 respectively. This performance
reduction is caused by the increasing number of emerging classes; 8 in Table~\ref{tab:result1}, 10 in
Table~\ref{tab:result2}, and 17 in Table~\ref{tab:result3}. Another reason is  
decreasing number of training instances;
42 in Table~\ref{tab:result1}, 27 in Table~\ref{tab:result2}, 
and 12 in Table~\ref{tab:result3}.
As can be seen in these Tables, our name disambiguation dataset contains a large number of emerging
records in the test data, all of these records will be misclassified with certainty by any traditional 
exhaustive name
disambiguation methods. This is our main motivation for designing a non-exhaustive classification
framework for online name entity disambiguation task.

Now we compare our method with the four competing methods. We observe that
our proposed Bayesian non-exhaustive classification method performs the best for 
11, 11, and 12 name references (out of 15) in Table~\ref{tab:result1}, Table~\ref{tab:result2},
and Table~\ref{tab:result3}, respectively. Besides, the margin of performance difference between
our method and the second best method is large, typically between 0.05 and 0.20 in
terms of macro-F1 score. 
For an example, consider the name entity Lei Wang in 
Table~\ref{tab:result2}, even though it has 67 emerging records with $45$ emerging classes, our method achieves
0.723 macro-F1 score for this name reference; whereas the second best method achieves only 0.560---which
is smaller by 0.163. The relatively good performance of the proposed method 
may be due to our non-exhaustive learning methodologies. It also suggests that the base distribution used by the proposed Dirichlet process prior model whose parameters are estimated using data from known classes can be generalized for the class distributions of unknown classes as well. 

In contrast, among all the competing methods, Qian's Method and Khabsa's Method perform the worst 
as they fail to incorporate prior information about class distribution into the models and the results 
are very sensitive to the selections of threshold parameters. On the other hand both BernouNaive-HAC 
and NNMF-SVM-HAC operate in an off-line framework. Although for some name references F1 scores obtained 
by these techniques are higher than our proposed method,  there is a clear trend favoring our proposed 
method over these methods---latter cannot explicitly identify streaming records of new ambiguous classes 
in an online setting.

In Table~\ref{tab:result4}, using the data records of most recent $2$ years as test set, we
present the results of automatic estimation of the number of distinct entities in test set. As shown 
in Table~\ref{tab:result4}, \#Actual Authors is the ground truth number of real-life persons among the records
in the test set, and \#Predicted Authors is the value predicted by our proposed method. We can see that the estimated numbers are close to the actual numbers for most name references.  For example, for the name reference of ``Bo Liu", our predicted result is exactly the same as the actual one. 
Overall these results demonstrate that our proposed framework offers a robust  approach to accurately estimate the number of actual real-life persons,  especially when the records appear in a streaming fashion. 

\begin{table}[t!]
\centering
\scalebox{0.80}{
\begin{tabular}{c c c}
\toprule
Name Reference & \# Actual Authors & \# Predicted Authors \\ \midrule
Kai Zhang & 13 & 10 \\
Bo Liu & 15  & 15 \\
Jing Zhang & 52 & 67 \\
Yong Chen & 10 & 12 \\
Yu Zhang & 45 & 37 \\
Hao Wang & 17 & 24 \\
Wei Xu & 18 & 20 \\
Lei Wang & 41 & 51 \\
Bin Li & 18 & 21 \\
Feng Liu & 16 & 23 \\
Lei Chen & 26 & 40 \\
Ning Zhang & 16 & 17 \\
David Brown & 7 & 6 \\
Yang Wang & 31 & 46 \\
Gang Chen & 25 & 47 \\
\bottomrule
\end{tabular}}
\caption{Results of number of predicted distinct real-life persons under our proposed method using records with most recent 2 years as test set}
\label{tab:result4}
\vspace{-0.10in}
\end{table}

\subsection{Feature Contribution Analysis}

\begin{figure}
\includegraphics[width = \linewidth]{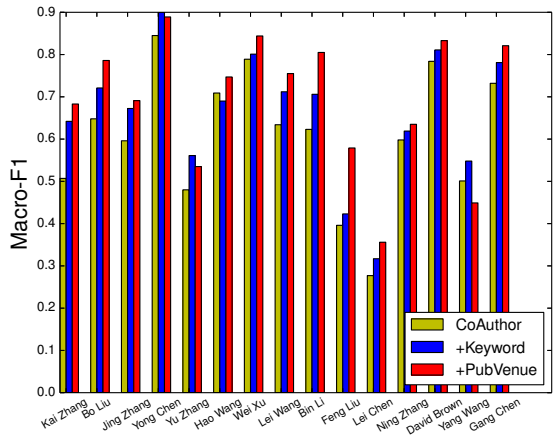}
\caption{Feature contribution analysis using most recent 2 years' publication records as test set}
\label{fig:feature-analysis}
\vspace{-0.2in}
\end{figure}

We also investigate the contribution of each of the defined features (coauthor, keyword, venue) for the task of online name disambiguation.
Specifically, we first rank the individual features by their performance in terms of Macro-F1 score, then add the features one by one in the order of their disambiguation power. In particular, we first use author-list, followed by Keywords, and Publication Venue. 
In each step, we evaluate the performance of our proposed online name disambiguation method using the
most recent two years' publication records as test set. Figure~\ref{fig:feature-analysis} shows the Macro-F1 value of our method with different feature combinations. As we can see from this figure, after adding each feature group we observe improvements for most of the name references. Similar patterns are observed when we use different number of publication records as test set. 

\vfill\eject
\subsection{Study of Running Time}

A very desirable feature of our proposed Bayesian non-exhaustive classification model is its running time. For example, using the most recent two years' records as test set, on the name reference ``Kai Zhang" containing $66$ papers with $10$ latent dimensionality, it takes around $0.29$ seconds on average to assign the test papers to different real-life authors for one-sweep Gibbs sampler. For the name reference ``Lei Wang" with $308$ papers using same number of latent dimensionality, it takes around $1.95$ seconds on average under the same setting. This suggests only a linear increase in computational time with respect to the number of records. However
in addition to number of records,
the computational time depends on other factors, such as the latent dimensionality and the number of classes generated, which in turn depends on the values of the hyperparameters used in the data model.

\section{Conclusion}

To conclude, in this paper we propose a Bayesian non-exhaustive classification framework for online name entity
disambiguation. Given sequentially observed record streams, our method classifies the incoming records 
into existing classes, as well as emerging classes by using one sweep Gibbs sampler
for learning posterior probability of a Dirichlet process mixture model. 
Our experimental results on bibliographic datasets show that the proposed method 
significantly outperforms the existing state-of-the-arts for disambiguating authors' name in online setting.

\bibliographystyle{abbrv}
\balance
\scriptsize{
\bibliography{online_disambiguation} 
}
\end{document}